\documentclass[pra,twocolumn,aps,english,pra,superscriptaddress,showpacs]{revtex4-1}
\usepackage{amsmath}
\usepackage{amssymb}
\usepackage{graphicx}
\usepackage{babel}
\usepackage{cases}
\usepackage{eufrak}

\makeatletter

\begin{document}

\title{Ground-state cooling of dispersively coupled optomechanical system in unresolved sideband regime via dissipatively coupled oscillator}

\author{Yu-Xiang Zhang}
\email{iyxz@mail.ustc.edu.cn}
\affiliation{Hefei National Laboratory for Physical Sciences at Microscale and Department of
Modern Physics, University of Science and Technology of China, Hefei, Anhui 230026, China}
\affiliation{The CAS Center for Excellence in QIQP and the Synergetic Innovation Center
for QIQP, University of Science and Technology of China, Hefei, Anhui 230026, China}
\affiliation{Kuang Yaming Honors School, Nanjing University, Nanjing, Jiangsu 210093, China}
\affiliation{Research Center of Integrative Molecular Systems, Institute for Molecular Science, National Institutes of Natural Sciences, 38 Nishigo-Naka, Myodaiji, Okazaki, Aichi 444-8585, Japan}
\author{Shengjun Wu}
\email{sjwu@nju.edu.cn}
\affiliation{Kuang Yaming Honors School, Nanjing University, Nanjing, Jiangsu 210093, China}
\author{Zeng-Bing Chen}
\email{zbchen@ustc.edu.cn}
\affiliation{Hefei National Laboratory for Physical Sciences at Microscale and Department of
Modern Physics, University of Science and Technology of China, Hefei, Anhui 230026, China}
\affiliation{The CAS Center for Excellence in QIQP and the Synergetic Innovation Center
for QIQP, University of Science and Technology of China, Hefei, Anhui 230026, China}
\author{Yutaka Shikano}
\email{yshikano@ims.ac.jp}
\affiliation{Research Center of Integrative Molecular Systems, Institute for Molecular Science, National Institutes of Natural Sciences, 38 Nishigo-Naka, Myodaiji, Okazaki, Aichi 444-8585, Japan}
\affiliation{Institute for Quantum Studies, Chapman University, 1 University Dr., Orange, CA 92866, USA}
\affiliation{Materials and Structures Laboratory, Tokyo Institute of Technology, 4259 Nagatsuta, Yokohama 226-8503, Japan}
\date{\today}

\begin{abstract}
    In the optomechanical cooling of a dispersively coupled oscillator,
    it is only possible to reach the oscillator ground state in the resolved sideband regime,
    where the cavity-mode line width is smaller than the resonant frequency of the mechanical
    oscillator being cooled. In this paper, we show that the dispersively coupled system can be cooled to the ground state in the unresolved
    sideband regime using an ancillary oscillator, which is coupled to the same optical mode via dissipative interaction.
    The ancillary oscillator has a resonant frequency close to that of the target oscillator; thus, the ancillary oscillator is also in the
    unresolved sideband regime. We require only a single blue-detuned laser mode to drive the cavity.
\end{abstract}
\maketitle

\section{Introduction}
Quantum optomechanics studies the interactions between photons and mechanical oscillators.
To facilitate applications yielding precise measurements and control features \cite{precise-1,precise-2,precise-3,precise-4,precise-5,precise-6},
certain quantum information processing techniques \cite{infor-1,infor-2,infor-3,infor-4,infor-5},
quantum foundations \cite{qc-1,qc-2,qc-3}, etc.,
it is necessary to prepare an oscillator in an almost pure state 
close to the zero-point vibration.
Therefore, techniques for ground-state cooling that remedy the effects of stochastic driving from the thermal
environment \cite{groundstate-1,groundstate-2,groundstate-3,GHz2,feedback-1,feedback-2,feedback-3,feedback-4,feedback-5}
are fundamentally important \cite{opto-1,opto-2,review,review-chen,review-chinese}.
However, as regards attempts to realize the ground state of such an oscillator, even the $\textsuperscript{3}{\rm He}/\textsuperscript{4}{\rm He}$
dilution refrigerator is insufficient,
unless the oscillator has very high resonant frequency (GHz) \cite{GHz1,GHz2}.
Thus, it is necessary to exploit laser cooling schemes (or microwave techniques, etc., depending on the system).

Lasers are employed in optomechanical cooling in two ways \cite{review}.
One application is to measure the instant position of the oscillator.
Then, an appropriate friction force is exerted in order to reduce the oscillation amplitude; this is known as cold damping quantum feedback \cite{feedback-1,feedback-2,feedback-3,feedback-4,feedback-5}, and 
the efficacy of this method depends on the measurement precision and the
feedback loop quality \cite{limit-A}.
However, the second application is more interesting to us. It is based on that fact that, in some parameter regimes,
the optomechanical interaction drives the cooling process automatically.
The most extensively investigated interaction to generate this passive (or self-) cooling phenomenon is
 {\it dispersive} coupling, which is named for the feature in which the mechanical oscillator displacement
changes the resonant frequency of the optical mode.
Ground-state cooling involving such scenarios is valid only
in the resolved sideband regime \cite{sideband-07-1,sideband-07-2,limit-A},
where the mechanical resonant frequency
is larger than the line width (or the damping rate) of the optical mode.
Thus, this phenomenon is also referred to as sideband cooling.

From a practical perspective, an unresolved sideband regime
allows one to use small drive detuning and, thus, small input power.
Almost all the realized optomechanical systems with heavier oscillator functionality
operate in the unresolved sideband regime \cite{review}, and a well-known example
is the suspended mirrors of the laser interferometer gravitational wave detector \cite{aligo}.
Therefore, eliminating the considerable limitation imposed by the requirement for a resolved sideband
for ground-state cooling would enrich the optomechanical toolbox in a meaningful way.

Note that sideband resolution is not a stringent requirement in optomechanical systems with
{\it dissipative} interaction, where the oscillator displacement changes the damping rate (or line width) of
the optical mode \cite{diss-1,diss-X,diss-expe,diss-limit,diss-realize,diss-strong,MS-ex,new-1,new-2}.
Although works on that topic are relatively rare, the predicted cooling has been verified experimentally \cite{MS-ex}
in an optomechanical system based on the Michelson-Sagnac interferometer \cite{diss-realize}.
Thus, the dissipative system is promising and merits further development. Here,
we show that this system also sheds light on the dispersive system problem, i.e.,
ground-state cooling in the unresolved sideband regime.

In this study, we demonstrate that the dissipatively coupled oscillator in a hybrid optomechanical system comprised of both dispersively and dissipatively coupled oscillators cools not only itself, but also the dispersively coupled oscillator, which
is coupled to the same cavity mode. Importantly, both of the oscillators are in
the unresolved sideband regime.
Thus, a solution to the difficult problem of
cooling the dispersively coupled oscillator is provided.
Note that this approach is not the first reported technique of this kind;
however, the existing schemes for ground-state cooling in the unresolved sideband, such as
cooling with optomechanically induced transparency (OMIT) \cite{eit-2,eit-cooling},
coupled-cavity configurations \cite{mt-cav-2015,mt-cav-2014},
atom-optomechanical hybrid systems \cite{atom-2015,atom-1,atom-2,atom-3,atom-4,atom-5,atom-6,atom-7},
and the recently proposed scheme using quantum non-demolition interactions \cite{QND},
require multiple driving lasers,
multiple optical modes, high-quality cavities, and ground-state atom ensembles.
Compared with those methods, our proposal offers a simpler option
for cases in which dissipative coupling is accessible.

The remainder of this paper is organized as follows. In Sec. \ref{intro}, we briefly introduce
the two kinds of optomechanical coupling and the quantum noise approach to the
cooling limit. In Sec. \ref{scheme}, we use the quantum noise approach to analyze
the possibility of ground-state cooling in our system, which requires
only one cavity mode and one laser driving mode.
In Sec. \ref{exact-section}, we use the exact solutions to the equations of motion
to confirm the achievement of ground-state cooling in the unresolved sideband regime.
In Sec. \ref{discussion}, we discuss and compare the existing schemes
and, in Sec. \ref{conclusion}, we present the conclusion. Note that the natural unit $\hbar = 1$
is used throughout this paper.

\section{Brief overview of passive cooling induced by quantum optical noise}\label{intro}
\begin{figure}[bt]
    \includegraphics[width=0.34\textwidth]{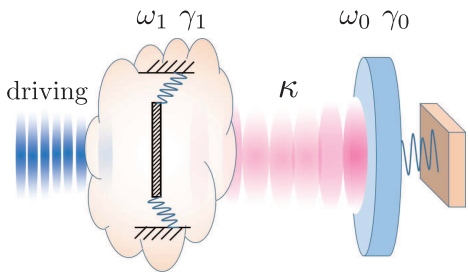}
    \caption{Hybrid optomechanical system of dispersively and
    dissipatively coupled oscillators. The oscillators
    are coupled to the same cavity mode, with line width $\kappa$. This
    mode is driven by one laser mode only, with blue detuning $\Delta>0$.}
    \label{system}
\end{figure}

Passive (or self-) cooling utilizes the optomechanical interaction to diminish the
oscillation amplitude. The system investigated in this paper is
illustrated in Fig. \ref{system}. It
consists of two oscillators, with one being dispersively coupled and the other being dissipatively
coupled to the same cavity mode. In this section, we introduce these two
kinds of optomechanical coupling separately.
The quantum noise approach to optomechanical cooling
is emphasized and exploited in the next section in order to evidence
the validity of our scheme.

\subsection{Cooling with dispersive coupling}\label{review-dispersive}
We first specify the definition of dispersive coupling.
In a typical cavity optomechanics setup, the cavity end mirror plays the role of the oscillator.
Displacement of this mirror alters the cavity length and, thus, the resonant frequency of the cavity mode;
this leads to {\it dispersive coupling}, which can be described by the Hamiltonian
\begin{equation}
H_{int,0}=-g_0(\hat{b}_0+\hat{b}_0^\dagger)(\hat{a}^\dagger\hat{a}-\langle \hat{a}^\dagger\hat{a}\rangle),
\label{dispersive coupling}
\end{equation}
where $g_0$ is the coupling strength, $\hat{b}_0$
is the phonon annihilation operator, and $\hat{a}$ annihilates the cavity-mode photons.
Hereafter, the suffix ``0'' will be reserved for this dispersively coupled oscillator.

In the expression for $H_{int,0}$, we have subtracted the steady-state photon number
$\langle\hat{a}^\dagger\hat{a}\rangle=\alpha^2_s$, following the formulae of Ref. \cite{sideband-07-1}.
It is assumed that $\alpha_s$ is a real number for convenience.
This subtraction frees us from redefining
parameters such as the cavity length and $\Delta$ (as in Ref. \cite{sideband-07-2}),
which are altered sightly because of the radiation pressure.

Suppose the cavity is driven by a single laser mode. The photons
in the cavity strive to reach a frequency close to resonance. Therefore, if the
input laser (with frequency $\omega_d$) is red detuned to the cavity mode ($\omega_c$),
which corresponds to the detuning $\Delta=\omega_d-\omega_c<0$, the input photons preferentially
extract energy from the oscillator via the interaction described by Eq. (\ref{dispersive coupling}).
This mechanism constitutes a simple explanation of dispersive cooling, from which we can
further infer that the optimal detuning is $\Delta=-\omega_0$ (the resonant frequency
of the oscillator) in the weak coupling regime.

It is convention to
refer to the energy quanta of the mechanical oscillator as phonons.
Then, the Fermi golden rule indicates that, for the oscillator,
the optics-induced emission or absorption of
phonons occurs at a rate that depends on the
the photon-number fluctuation spectrum at $\omega=\pm \omega_0$.
This spectrum is given by
\begin{equation}
\begin{aligned}
S^{0}_{nn}[\omega]&=\int_{-\infty}^{\infty}d\tau e^{i\omega\tau}\langle\delta\hat{n}(\tau)\delta\hat{n}(0)\rangle,\\
&=\frac{\kappa \alpha_s^2}{\kappa^2/4+(\omega+\Delta)^2},
\label{0-spectrum}
\end{aligned}
\end{equation}
where $\delta\hat{n}=\hat{a}^\dagger\hat{a}-\langle\hat{a}^\dagger\hat{a}\rangle$ and $\kappa$
is the line-width of the cavity mode in question.
The damping rate induced by the optics is then
\begin{equation}
\gamma_{opt,0}=g_0^2(S_{nn}^0[\omega_0]-S^0_{nn}[-\omega_0]).
\end{equation}
The balance between the optics-induced emission and absorption leads to
a steady-state phonon occupation (when $\Delta <0$)
\begin{equation}
\begin{aligned}
n_{opt,0}=&\frac{S^0_{nn}[-\omega_0]}{S^0_{nn}[\omega_0]-S^0_{nn}[-\omega_0]},\\
         =&-\frac{(\omega_0+\Delta)^2+\kappa^2/4}{4\omega_0\Delta}.
         \end{aligned}
         \label{n-opt-0}
\end{equation}
Incorporating the influence of the thermal environment,
the full expression of the phonon number is
\begin{equation}
\tilde{n}_0=\frac{\gamma_{opt,0}n_{opt,0}+\gamma_0 n_{th,0}}{\gamma_{opt,0}+\gamma_0},
\label{cooling-limit-noise}
\end{equation}
where $\gamma_0$ and $n_{th,0}$ represent the damping rate and the phonon
occupation in the absence of the optical system, respectively.

Equation (\ref{cooling-limit-noise}) implies that ground-state cooling occurs
if $\gamma_{opt,0}\gg\gamma_0$ and $n_{opt,0}\ll 1$.
In the resolved sideband regime, where $\omega_0>\kappa$, $n_{opt,0}$ reaches
its minimum $(\frac{\kappa}{4\omega_0})^2$ when $\Delta=-\omega_0$.
Ground-state cooling is, therefore, possible in this regime.
In contrast, in the unresolved sideband regime, where $\kappa\gg\omega_0$, the minimum
of $n_{opt,0}$ is $\frac{\kappa}{4\omega_0}$, for which the detuning should be $\Delta=-\kappa/2$.
Obviously, it is difficult for $n_{opt}$ to be close to zero, and this leads
to problematic cooling of
the dispersive optomechanical system in the unresolved sideband regime.

\subsection{Cooling with dissipative coupling}\label{intro-diss}
If the displacement of the oscillator modifies
$\kappa$, the induced coupling is called {\it dissipative}.
Dissipative coupling was first proposed in Ref. \cite{diss-1} and then experimentally implemented in a
nanomechanical system \cite{diss-expe,diss-X}.
Reference \cite{diss-realize} gave another experimental construction involving
the Michelson-Sagnac interferometer, which was implemented in experiment very recently \cite{MS-ex}.
The strong coupling effects \cite{diss-strong}, cooling limit \cite{diss-limit},
anomalous dynamic backactions \cite{new-1}, and stabilities \cite{new-2} associated with dissipative cooling
have also been studied recently. However, research on this topic is rare in comparison with
the literature on dispersive optomechanics.

Throughout this paper, terms related to the dissipatively coupled oscillator
are marked by the subscript ``1''.
The dissipative coupling is described by the Hamiltonian
\begin{equation}
H_{int,1}=-i g_1 \sqrt{\frac{\kappa}{2\pi\rho}}\sum_{q}(\hat{a}^\dagger\hat{b}_q
  -\hat{b}^\dagger_q\hat{a})(\hat{b}^\dagger_1+\hat{b}_1),
  \label{dissi-couple}
\end{equation}
where the summation is taken over the environment modes $\hat{b}_q$,
the density of states of which is assumed to be a constant $\rho$.

The cooling mechanism is less transparent than in the dispersive case.
Following the quantum noise approach, the force spectrum responsible
for the optomechanical cooling is
\begin{equation}
S_{FF}[\omega]=\kappa g_1^2\alpha_s^2\frac{(\omega+2\Delta)^2}{\kappa^2/4+(\omega+\Delta)^2}.
\label{f-spectrum}
\end{equation}
The rate of phonon generation is $S_{FF}[-\omega_1]$ and
that of phonon loss is $S_{FF}[\omega_1]$.
The optically induced damping rate is, therefore, given by their difference
\begin{equation}
\gamma_{opt,1}=S_{FF}[\omega_1]-S_{FF}[-\omega_1].
\end{equation}
The further formulas for $n_{opt,1}$ and the final phonon number
$\tilde{n}_1$ are in the same forms as Eqs. (\ref{n-opt-0}) and (\ref{cooling-limit-noise}).

Despite the sideband resolution, both $S_{FF}[-\omega_1]$ and
$n_{opt,1}$ vanish if the driving laser is blue detuned at $\Delta=\omega_1 /2$.
The conclusion can then be drawn that ground-state cooling is robust, provided a high mechanical quality factor
and proper pre-cooling is supplied.
This point is verified by the numerical calculations given in
\cite{diss-1,diss-strong}, which show that $\tilde{n}_1$ can be reduced to the level of $10^{-2}$
from the initial occupation $n_{th,1}=100$.

\section{Ground-state cooling: Evidence from quantum noise approach}\label{scheme}
In this section, we use the quantum noise approach to
show that the dispersively coupled oscillator can be cooled to
the ground state with the assistance of a dissipative
oscillator. Both of these oscillators are in the unresolved sideband regime.
As previously, the quantities of the target dispersive oscillator
are marked with subscript ``0'', and those of the
ancillary dissipative oscillator are indicated by subscript ``1''.

\subsection{Model and equations}
In the frame of the driving laser, the Hamiltonian of the entire system is
\begin{equation}
\begin{aligned}
H=&-\Delta (\hat{a}^\dagger\hat{a}-\langle \hat{a}^\dagger\hat{a}\rangle) + \omega_0\hat{b}_0^\dagger\hat{b}_0+\omega_1\hat{b}_1^\dagger\hat{b}_1\\
& +H_{int,0}+H_{int,1}+H_{\kappa}+H_{\gamma_0}+H_{\gamma_1},
\end{aligned}
\end{equation}
where $H_{\gamma_0}$ and $H_{\gamma_1}$ are the
thermal damping terms of the two oscillators. Further,
\begin{equation}
H_{\kappa}=-i\sqrt{\frac{\kappa}{2\pi\rho}}\sum_{q}(\hat{a}^\dagger\hat{b}_q
  -\hat{b}^\dagger_q\hat{a}),
\end{equation}
describes the damping of the cavity mode induced by the environment, including
the driving mode. The free Hamiltonian of the environment modes is not presented explicitly.

To derive the equations of motion, the input-output formalism \cite{review-noise} is considered, which yields
\begin{equation}
\sqrt{\frac{\kappa}{2\pi\rho}}\sum_{q} \hat{b}_q=\sqrt{\kappa}\hat{a}_{in}
+\frac{\kappa}{2}\hat{a}+g_1\frac{\kappa}{2}(\hat{b}_1+\hat{b}^\dagger_1)\hat{a}.
\end{equation}
This formula can be used to obtain the quantum Langevin equations and, also, to
further linearize them by separating the operators associated
with optics into steady-state and quantum-fluctuation components.
That is, we assume that $\alpha_s$ is sufficiently significant that
$\hat{a}_{in}=\bar{a}_{in}+\hat{d}_{in}$ and $\hat{a}=\alpha_s+\hat{d}$.
The two steady-state average amplitudes are correlated by
$(i\Delta-\kappa/2)\alpha_s=\sqrt{\kappa}\bar{a}_{in}$.

After the linearizion, the quantum Langevin equations of the cavity mode and the mechanical
oscillators are expressed as
\begin{eqnarray}
 \nonumber
 \dot{\hat{d}} &=& (i\Delta-\frac{\kappa}{2})\hat{d}-
  \sqrt{\kappa}\hat{d}_{in}+ig_0\alpha_s(\hat{b}_0+\hat{b}_0^\dagger) \\
  & &-g_1\alpha_s(i\Delta+\frac{\kappa}{2})(\hat{b}_1+\hat{b}_1^\dagger),
  \label{equ-d}\\
\dot{\hat{b}}_0 &=& -(i\omega_0+\frac{\gamma_0}{2})\hat{b}_0-
  \sqrt{\gamma_0}\hat{b}_{in,0}+ig_0\alpha_s(\hat{d}+\hat{d}^\dagger),
  \label{equ-b0}\\
\nonumber\dot{\hat{b}}_1 &=& -(i\omega_1+\frac{\gamma_1}{2})\hat{b}_1-
  \sqrt{\gamma_1}\hat{b}_{in,1}-g_1\alpha_s\sqrt{\kappa}(\hat{d}_{in}-\hat{d}_{in}^\dagger)\\
   & &-ig_1\alpha_s\Delta(\hat{d}+\hat{d}^\dagger)
    -g_1\alpha_s\frac{\kappa}{2}(\hat{d}-\hat{d}^\dagger).
    \label{equ-b1}
\end{eqnarray}
In those equations, the noise terms with subscript ``in" satisfy the correlations
\begin{equation}
\begin{aligned}
\langle \hat{d}_{in}(t)\hat{d}^\dagger_{in}(t^\prime) \rangle & =\delta(t-t^\prime), \\
\langle\hat{b}_{in,k}(t)\hat{b}^\dagger_{in,k}(t^\prime)\rangle &= (n_{th,k}+1)\delta(t-t^\prime),\\
\langle\hat{b}^\dagger_{in,k}(t)\hat{b}_{in,k}(t^\prime)\rangle &= n_{th,k}\delta(t-t^\prime),
\end{aligned}
\label{correlation}
\end{equation}
where $n_{th,k}$ is the thermal equilibrium phonon number of oscillator $k$ (where $k = 0, 1$).
All other two-operator correlation functions vanish.

We can include the conjugate equations and write the complete set of equations in the form
$\frac{d}{dt}\vec{V}=M\cdot \vec{V}+\vec{V}_{in}$, where
$\vec{V}=(\hat{d},\hat{d}^\dagger,\hat{b}_0,\hat{b}_0^\dagger,\hat{b}_1,\hat{b}_1^\dagger)$ and
$\vec{V}_{in}$ has a similar definition.
The system is dynamically stable if the eigenvalues
of the matrix $M$ have no negative real parts, which can be easily checked after
the parameters are fixed.
We find that the issue of stability forbids
$\Delta=-\omega_1$
in the unresolved sideband regime when $\kappa\gg\omega_1$, while
this detuning allows ground-state cooling when $\kappa$ and $\omega_1$ are comparable
or in the resolved sideband regime \cite{,diss-1,diss-strong}.

\subsection{Photon-number fluctuation spectrum}
In Sec. \ref{review-dispersive}, it was shown that
the photon-number fluctuation spectrum
is responsible for the cooling of oscillator 0.
To calculate this spectrum, we follow the strategy used in Ref. \cite{eit-cooling}, i.e.,
we ignore the backaction of oscillator 0 to the optical field
but consider that of oscillator 1.
Note that the validity of this strategy is examined in Sec. \ref{exact-section} via a numerical calculation of
the cooling limit from the linearized Langevin equations.

Applying Fourier transformations to Eqs. (\ref{equ-d}) and (\ref{equ-b1})
with $g_0=0$, we obtain a series of algebraic equations that are exactly solvable.
The Fourier transformation is conducted with the conventions
\begin{equation}
\hat{f}_\omega=\int_{-\infty}^{\infty}dt e^{i\omega t}\hat{f}(t),
\end{equation}
and $\hat{f}^\dagger_\omega\equiv(\hat{f}^\dagger)_\omega=(\hat{f}_{-\omega})^\dagger$.
The photon number fluctuation annihilation operator $\hat{d}(t)$ is transformed to
\begin{equation}
\hat{d}_\omega=-\sqrt{\kappa}\chi_{c,\omega}\hat{d}_{in,\omega}-
g_1\alpha_s(i\Delta+\frac{\kappa}{2})\chi_{c,\omega}\hat{x}_{1,\omega},
\label{d-w}
\end{equation}
where $\hat{x}_{1,\omega}=\hat{b}_{1,\omega}+\hat{b}_{1,\omega}^\dagger$ and
\begin{equation}
\begin{aligned}
\hat{b}_{1,\omega}=&-\frac{\sqrt{\gamma_1}}{N[\omega]}\{\chi_{1,-\omega}^{*-1}\hat{b}_{in,1,\omega}
-i\Sigma[\omega](\hat{b}_{in,1,\omega}+\hat{b}_{in,1,\omega}^\dagger)\}\\
 & -g_1\alpha_s \frac{\sqrt{\kappa}}{N[\omega]}\chi^{*-1}_{1,-\omega}
 [\alpha(\omega)\hat{d}_{in,\omega}-\alpha^{*}(-\omega)\hat{d}^\dagger_{in,\omega}].
 \label{x-1}
\end{aligned}
\end{equation}
In the above equations, the susceptibilities of the optical mode $\chi_c$
and oscillator 1 $\chi_1$ are defined as
\begin{eqnarray}
\chi^{-1}_{c,\omega} &=& \kappa/2-i(\omega+\Delta),\\
\chi^{-1}_{1,\omega} &=& \gamma_1/2-i(\omega-\omega_1),
\end{eqnarray}
and the other quantities are
\begin{eqnarray}
\alpha(\omega) &=& 1-\chi_{c,\omega}(i\Delta+\kappa/2), \\
N[\omega] &=& \chi^{-1}_{1,\omega}\chi^{*-1}_{1,-\omega}+2\omega_1\Sigma[\omega], \label{near}
\end{eqnarray}
with the ``self-energy'' term
\begin{equation}
\Sigma[\omega]=ig_1^2\alpha_s^2 \left\{ \chi_{c,\omega} \left( i\Delta+\frac{\kappa}{2} \right)^2-\chi^{*}_{c,-\omega} \left( i\Delta-\frac{\kappa}{2} \right)^2 \right\}.
\label{self-energy}
\end{equation}

Next, we proceed to examine the photon-number fluctuation spectrum.
As $\delta\hat{n}=\alpha_s(\hat{d}+\hat{d}^\dagger)$ and
\begin{equation}
\begin{aligned}
\hat{d}_\omega+\hat{d}^\dagger_{\omega}=&-\sqrt{\kappa}(\chi_{c,\omega}\hat{d}_{in,\omega}+
         \chi^{*}_{c,-\omega}\hat{d}^\dagger_{in,\omega})\\
         &-g_1\alpha_s A(\omega)\hat{x}_{1,\omega},
\end{aligned}
\label{photon-noise}
\end{equation}
where $A(\omega)=\chi_{c,\omega}(i\Delta+\kappa/2)+\chi^*_{c,-\omega}(-i\Delta+\kappa/2)$,
the fluctuation spectrum can be expressed as
\begin{equation}
\begin{aligned}
S_{nn}[\omega]=& S_{nn}^0[\omega]+g_1^2\alpha_s^4|A(\omega)|^2 S^1_{xx}[\omega]\\
& +\sqrt{\kappa}g_1\alpha_s^3\{A(-\omega)S_{cx}[\omega]+A(\omega)S_{xc}[\omega]\}.
\end{aligned}
\label{all-spectrum}
\end{equation}
Therein, $S_{nn}^0$ is the bare spectrum given in Eq. (\ref{0-spectrum}) and $S_{xx}^1[\omega]$ is the
spectrum of the position of oscillator 1
\begin{equation}
S_{xx}^1[\omega]=\int_{-\infty}^{\infty}\frac{d\omega'}{2\pi}\langle\hat{x}_{1,\omega}\hat{x}_{1,\omega'}\rangle.
\end{equation}
$S_{cx}[\omega]$ is the correlation between oscillator 1 and the
$X$-quadrature of the optical mode in the absence of oscillator 1, such that
\begin{equation}
S_{cx}[\omega]=\int_{-\infty}^{\infty}\frac{d\omega'}{2\pi}\langle \hat{M}_c[\omega]\hat{x}_{1,\omega'}\rangle,
\label{interference}
\end{equation}
where $\hat{M}_c[\omega]=\chi_{c,\omega}\hat{d}_{in,\omega}+\chi^*_{c,-\omega}\hat{d}^\dagger_{in,\omega}$.
$S_{xc}$ is defined similarly.

\subsection{Ground-state cooling}\label{ground-resonance}
We know that, in the unresolved
sideband regime, the gap between $S^0_{nn}[\pm\omega_0]$ is too narrow
to support ground-state cooling.
Now, the photon number fluctuation spectrum [Eq. (\ref{all-spectrum})] contains
two more terms than the bare $S^0_{nn}$.
Thus, we examine the manner in which these additional terms unbalance
the spectrum. For this purpose, the exact expressions presented
in the previous subsection are
not sufficiently transparent for the result to be observed.
Therefore, we further simply them by assuming weak coupling.

\subsubsection{$S_{nn}[\omega]$ shape}
When oscillator 1 is decoupled to the optical field,
its position-position correlation spectrum is expressed in terms of the ``bare''
damping rate $\gamma_1$, resonant frequency (incorporated in the susceptibility $\chi_{1,\omega}$), and thermal equilibrium phonon occupation $n_{th,1}$: $$\gamma_1\{|\chi_{1,\omega}|^2(n_{th,1}+1)+|\chi_{1,-\omega}|^2 n_{th,1} \}.$$
When the opto-mechanical coupling is weak, the
effective damping rate $\tilde{\gamma}_1$, resonant frequency $\tilde{\omega}_1$, and phonon occupation
$\tilde{n}_1$ can be well defined.
By replacing the bare terms with the effective terms, $\chi_{1,\omega}$ is modified to
\begin{equation}
\tilde{\chi}_{1,\omega}=\frac{1}{\tilde{\gamma}_1/2-i(\omega-\tilde{\omega}_1)},
\label{eff-susce}
\end{equation}
where $\tilde{\gamma}_1=\gamma_1+\gamma_{opt,1}$. Further,
\begin{eqnarray}
\tilde{\omega}_1 &=& \omega_1+ \Re (\Sigma[\omega_1]) \approx \omega_1- 6g_1^2\alpha_s^2 \Delta,
\label{tilde-w1}\\
\tilde{\gamma}_1 &=& \gamma_1-2 \Im (\Sigma[\omega_1]) \approx \gamma_1 + 16g_1^2\alpha_s^2\Delta^2/\kappa,
\label{tilde-2}
\end{eqnarray}
where $\Re$ and $\Im$ represent the real and imaginary parts, respectively.
The approximations on the right hand side are conducted in the unresolved sideband regime, with the additional
assumption that $\Delta$ has the same scale as
$\omega_{0/1}$.
Additionally, for Eq. (\ref{tilde-w1}), a superior approximation is
\begin{equation}
\tilde{\omega}_1=\sqrt{\omega_1^2+2\omega_1\Re(\Sigma[\omega_1])}.
\label{precise-w1}
\end{equation}
The $\tilde{n}_1$ in Eq. (\ref{eff-susce})
corresponds to the phonon number obtained in the dissipative optomechanical system.
As discussed in Sec. \ref{intro-diss}, this can be very close to zero when $\Delta=\omega_1/2$.

Finally, the position spectrum is expressed as
\begin{equation}
S^1_{xx}[\omega]=\tilde{\gamma}_1\{|\tilde{\chi}_{1,\omega}|^2(\tilde{n}_1+1)
      +|\tilde{\chi}_{1,-\omega}|^2\tilde{n}_1 \},
\label{Sxx}
\end{equation}
where $\tilde{n}_1$ is the phonon occupation after the optomechanical cooling.
As explained in Sec. \ref{intro-diss},
$\tilde{n}_1$ approaches zero if $\Delta$ is properly
adjusted.
The interference terms between the displacement of oscillator 1
and the optics field are expressed as
\begin{equation}
\begin{aligned}
& S_{cx}[\omega]=-i2\omega_1g_1\alpha_s\frac{\sqrt{\kappa}}{N[-\omega]}\chi_{c,\omega}\alpha^*(\omega),\\
& S_{xc}[\omega]=i2\omega_1 g_1\alpha_s\frac{\sqrt{\kappa}}{N[\omega]}\chi^*_{c,\omega}\alpha(\omega).
\end{aligned}
\end{equation}
Note that the two interference terms are complex conjugate, as $N[\omega]=N^*[-\omega]$.
In particular, we can approximate
$N[\omega]\approx\tilde{\chi}^{-1}_{1,\omega}\tilde{\chi}^{*-1}_{1,-\omega}$.

Next, let us examine the shape of $S_{xx}^1[\omega]$.
Equation (\ref{Sxx}) indicates that there are two peaks at $\omega=\pm \tilde{\omega}_1$.
The ratio of their heights is $1+1/\tilde{n}_1$,
which implies an
extreme asymmetry in the limit $\tilde{n}_1\rightarrow 0$.
We wish to investigate whether or not this asymmetry can
meet our requirements for ground-state cooling.

\subsubsection{Cooling: Near the limit $\omega_0=\tilde{\omega}_1$}

If $|\omega_0-\tilde{\omega}_1|<\tilde{\gamma}_1$, $\pm \omega_0$ is located in the peak range.
We focus on the limit case at resonance in what follows, i.e.,
where $\omega_0=\tilde{\omega}_1$ or, more loosely, $|\omega_0-\tilde{\omega}_1|\ll\tilde{\gamma}_1$.
Other possibilities are studied in the next subsection.
\begin{figure}[b]
    \includegraphics[width=0.43\textwidth]{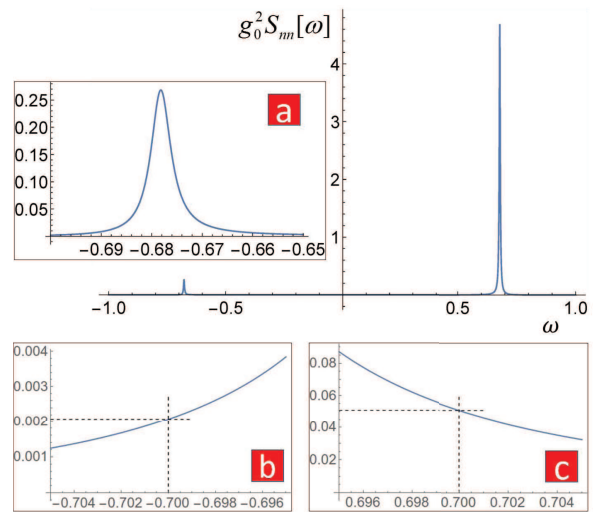}
    \caption{Photon-number fluctuation spectrum $g_0^2 S_{nn}[\omega]$. $\omega_1$ is set as the unit, which means $\omega_1=1$. The other parameters are: $\kappa=300$, $\gamma_1=10^{-6}$, $g_0\alpha_s=0.1$,
    $g_1\alpha_s=0.3$, $n_{1,th}=100$, and $\Delta=0.5$.
    The peaks are located at $\pm\tilde{\omega}_1\approx \pm 0.678$. The left peak is significantly shorter than the right. (a) Magnification of
    $-\tilde{\omega}_1\approx -0.678$ peak. If $\omega_0=\tilde{\omega}_1$, $n_{0,opt}\approx 0.061$.
    (b, c) If $\pm \omega_0$ is located away from the peaks, say at $\pm 0.7 $,
    then $g_0^2 S_{nn}[\omega_0]=0.051$ and $g_0^2S_{nn}[-\omega_0]=0.0021$, such that $n_{0,opt}\approx 0.042$. }
    \label{fig2}
\end{figure}

In the unresolved sideband regime and the considered case when $\omega$ is close to $\pm \omega_0$,
we have the approximations
\begin{equation}
\begin{aligned}
& \alpha(\omega)\approx\frac{\omega+2\Delta}{i\kappa/2},\qquad
A(\omega) \approx 2- i \frac{4 \omega}{\kappa},\\
& \chi_{1,\omega_0}=\frac{2}{\tilde{\gamma_1}},\qquad \chi_{c,\omega}\approx\frac{2}{\kappa}(1+\frac{2i(\omega+\Delta)}{\kappa}).
\end{aligned}
\label{approximations}
\end{equation}
We also assume that the mechanical oscillators have high quality factors and that
$\tilde{\gamma}_1\ll \tilde{\omega}_1$.
For convenience in what follows, we introduce a factor $\beta$ such that
$\omega_0+\tilde{\omega}_1=\beta^{-1}\omega_1$.
In fact, $\beta$ should be slightly larger than $1/2$.
Then, in $S_{cx}$ and $S_{xc}$, we have
\begin{numcases}
{\frac{\omega_1}{N[\omega]}=}
 i\beta \tilde{\chi}_{1,\omega}, & $\omega \sim \omega_0$, \\
 -i\beta \tilde{\chi}^{*}_{1,-\omega}, & $\omega\sim -\omega_0 $.
\end{numcases}
Following these preparations, the fluctuation spectrum at $\pm \omega_0$
can be expressed approximately as
\begin{eqnarray}
\nonumber
S_{nn}[\omega_0] &= & S_{nn}^{0}[\omega_0]+4 g_1^2\alpha_s^4 \tilde{\gamma}_1
\left[ \frac{4}{\tilde{\gamma}_1^2}(\tilde{n}_1+1)+\frac{\beta^2}{\omega_1^2}\tilde{n}_1 \right]\\
  & & +128 g_1^2\alpha_s^4 \beta \frac{(\Delta+2\omega_0)(\omega_0+2\Delta)}{\kappa^2\tilde{\gamma}_1},\\
\nonumber
S_{nn}[-\omega_0] &=& S_{nn}^0[-\omega_0]+ 4 g_1^2\alpha_s^4 \tilde{\gamma}_1
\left[ \frac{4}{\tilde{\gamma}_1^2}\tilde{n}_1+\frac{\beta^2}{\omega_1^2}(\tilde{n}_1+1) \right] \\
  & & -128 g_1^2\alpha_s^4 \beta \frac{(\Delta-2\omega_0)(2\Delta -\omega_0)}{\kappa^2\tilde{\gamma}_1},
  \label{resonance-w0}
\end{eqnarray}
Here, the conditions
$\kappa\gg \omega_0\gg \tilde{\gamma}_1\gg\gamma_1$ and $\tilde{\gamma}_1\kappa\sim \omega_1^2$,
which is supported by Eq. (\ref{tilde-2}), are adopted.
Then, the steady-state phonon number $n_{opt,0}$ is
\begin{equation}
n_{opt,0} \approx \tilde{n}_1 + O(\frac{\omega_0^2}{\kappa^2}) +
O(\frac{\tilde{\gamma}_1^2}{4\omega_1^2}).
\end{equation}
In the above equation, we have neglected terms of $O(\tilde{n}_1/\kappa^2)$ or higher.
Therefore, roughly speaking, $n_{opt,0}$ equates to $\tilde{n}_1$.
As the ancillary oscillator can be cooled to near the
ground state when $\Delta=\omega_1/2$,
ground-state cooling of the dispersively coupled oscillator 1
is indicated.

Figure \ref{fig2} clearly illustrates the asymmetrical feature of $S_{nn}[\omega]$.
The parameters are listed in the caption.
With those parameters, the standard sideband cooling with
the optimal setting $\Delta=-\kappa/2$ yields $n_{opt,0}\approx110.6>n_{th,0}=100$.
Beginning at this limit, the ancillary oscillator helps to unbalance the $S_{nn}[\omega]$ peaks and implies a new
limit at ${n}_{opt,0}=0.061$.

\subsubsection{Cooling: Off the peaks}
In this subsection, we release the restriction that
$|\omega_0-\tilde{\omega}_1|<\tilde{\gamma}_1$. However, the preliminary requirement that
the $\pm\omega_0$ are not far from the peaks is retained.
(Otherwise, the incorporation of oscillator 1 becomes meaningless.)

Now, the imbalance is no longer manifest from the shape of
$S_{nn}[\omega]$ and a detailed calculation is necessary.
To specify the conditions of this discussion, we assume that
$\tilde{\gamma}_1 \ll |\omega_0-\tilde{\omega}_1 | \ll \omega_0+\tilde{\omega}_1$ and $\Delta$ is on the scale
of $\omega_{0/1}$.
Hence, the approximations in Eq. (\ref{approximations})
remain valid, apart from the effective susceptibility of oscillator 1,
which should be altered to
\begin{equation}
\tilde{\chi}_{1,\pm \omega_0}\approx \frac{i}{\pm \omega_0-\tilde{\omega}_1}.
\end{equation}
We do not give the expressions of the
phonon-number fluctuation spectrum at $\pm \omega_0$ explicitly here.
In fact, all the terms exactly or approximately have
a factor $1/\kappa$.
However, among them, the dominant terms of $O[(\frac{\omega_0}{\omega_0-\tilde{\omega}_1})^2]$ are
\begin{eqnarray}
\nonumber
S_{nn}[\omega_0]& \approx & 4 g_1^2\alpha_s^4\tilde{\gamma}_1
 \frac{\tilde{n}_1+1}{(\omega_0-\tilde{\omega}_1)^2}, \\
S_{nn}[-\omega_0]& \approx & 4 g_1^2\alpha_s^4\tilde{\gamma}_1 \frac{\tilde{n}_1}{(\omega_0-\tilde{\omega}_1)^2}.
\label{spectrum-2case}
\end{eqnarray}
Besides the considerable
value of $g_1\alpha_s$ and the condition that $\frac{\omega_{0/1}}{|\omega_0-\tilde{\omega}_1|}\gg 1$,
the validity of the above approximations also relies on the preliminary that
$\frac{\tilde{n}_1}{(\omega_0-\tilde{\omega}_1)^2}$ covers $\frac{1}{(\omega_0+\tilde{\omega}_1)^2}$.
Then, we have
\begin{equation}
\gamma_{opt,0}= 4g_0^2 g_1^2\alpha_s^4
\frac{\tilde{\gamma}_1}{(\omega_0-\tilde{\omega}_1)^2},\qquad
n_{opt,0}\approx \tilde{n}_1.
\label{opt-2case}
\end{equation}
Therefore, building on the conditions that support Eq. (\ref{opt-2case}),
the dispersively coupled oscillator can be cooled to the ground state.

In Fig. \ref{fig2}, the roots of the
left and right peaks are magnified and their height imbalance is shown.
A cooling limit at $n_{opt,0}=0.042$ is predicted.

\subsection{Fixing the experimental parameters}

From the above analysis,
it seems that the optimal detuning is $\omega_1/2$, because this value yields the
minimal $\tilde{n}_1$. However, this intuition is not exactly correct.
Here, we inspect the cooling mechanism more closely and study
the parameters of the experimental variables that most strongly benefit cooling
via this quantum noise approach.

\subsubsection{Sources contributing to $S_{nn}[\pm \omega_0]$}

The photon-number fluctuation spectrum
is calculated from the $X$-quadrature $d_\omega+d^\dagger_\omega$,
which can be rewritten
according to the independent contributing sources as
\begin{equation}
\begin{aligned}
&-\sqrt{\kappa} \left\{
\chi_{c,\omega}+i \frac{2\omega_1 g_1^2\alpha_s^2\alpha(\omega)A(\omega)}{N[\omega]} \right\} \hat{d}_{in,\omega}\\
& \qquad +g_1 \alpha_s \sqrt{\gamma_1}\frac{A(\omega)}{N[\omega]}\chi^{*-1}_{1,-\omega}\hat{b}_{1,in,\omega}+ h.c.,
\label{2-sources}
\end{aligned}
\end{equation}
where $h.c.$ represents the hermitian conjugate, followed by the transformation $\omega\rightarrow -\omega$.
We remind readers that $A(\omega)=A^*(-\omega)$ and $N[\omega]=N^*[-\omega]$.

In Eq. (\ref{2-sources}), the second line is due to the thermal environment of
oscillator 1. Using Eq. (\ref{correlation}), it is apparent that, when
$n_{1,th}\gg 1$,
the thermal noise contributes to $S_{nn}[\pm \omega]$ equally (see the blue dashed lines in Fig. \ref{fig3}) with the value
\begin{equation}
\qquad g_0^2 g_1^2 \alpha_s^4 \gamma_1 |\frac{A(\omega)}{N[\omega]}|^2
(|\chi^{-1}_{1,\omega}|^2+|\chi^{-1}_{1,-\omega}|^2) n_{1,th}.
\label{thermal-contri}
\end{equation}
This implies that the imbalance in $S_{nn}[\omega]$ should be
attributed to the terms in the first line of Eq. (\ref{2-sources}),
i.e., those from the optical field.

The optical contribution component is interference between that
filtered by the cavity (represented by $\chi_{c,\omega}$) and that
introduced by oscillator 1 (represented by the $g_1$ factor).
Note that, for the former in the unresolved sideband regime, we have
$\chi_{c,\omega}\approx 2/\kappa$, which is a small constant that affects
both peaks equally.
Thus, the prospect of cooling is dependent on the latter.
This contribution must outweigh Eq. (\ref{thermal-contri}) by a large extent at $\omega=\omega_0$,
and take advantage of the destructive interference at $\omega=-\omega_0$.
Otherwise, the $S_{nn}[\omega]$ imbalance necessary for ground-state cooling cannot be established.

\subsubsection{Mechanical quality factor}

\begin{figure}[b]
    \includegraphics[width=0.47\textwidth]{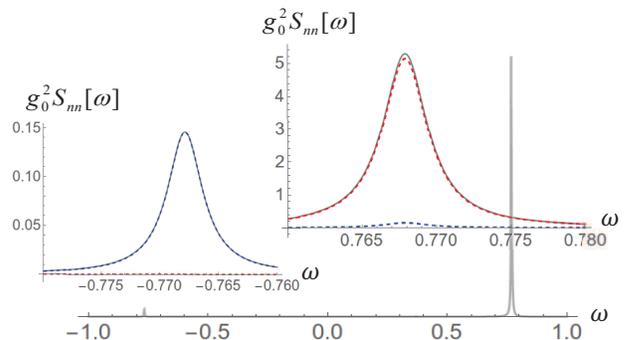}
    \caption{$g_0^2 S_{nn}[\omega]$ when $\Delta=\omega_0/2$. We set $\omega_1$ as the unit,
    which means $\omega_1=1$. The other experimental
    parameters are: $\kappa=300$ and $\gamma_1=10^{-6}$ (such that the quality factor of oscillator 1 is $10^6$),
    $g_0\alpha_s=0.1$, $g_1\alpha_s=0.3$, $n_{1,th}=100$, $\omega_0=0.76$, and $\Delta=0.38$.
    The peaks are located at $\pm\tilde{\omega}_1\approx \pm 0.768$. The peaks are presented separately.
    The red and blue dashed lines indicate the contributions from the optical field noise and the thermal environment of oscillator 1, respectively.
    $S_{nn}[\omega_0]=0.2814$ and $S_{nn}[-\omega_0]=0.0076$, such that $n_{0,opt}\approx 0.0277$. }
    \label{fig3}
\end{figure}

To achieve ground-state cooling, the contribution to $S_{nn}[\omega_0]$
from the first line of Eq. (\ref{2-sources}) 
must be significantly larger than that given in Eq. (\ref{thermal-contri}).
The ratio is
\begin{equation}
\frac{\omega_1/\gamma_1}{\kappa/\omega_1}
\frac{16 g_1^2 \alpha_s^2 (\omega+2\Delta)^2}
{ n_{th,1} (|\chi_{1,\omega}^{-1}|^2+|\chi_{1,-\omega}^{-1}|)},
\label{ratio}
\end{equation}
where we have approximated $\alpha(\omega)$ according to Eq. (\ref{approximations}).
As $\Delta$ is chosen to be of the same scale as $\omega_{0/1}$, the
value of the ratio depends on the first factor, i.e., 
the quality factor of oscillator 1, $\omega_1/\gamma_1$, vs. the sideband parameter $\kappa/\omega_1$.
Because of our unresolved sideband regime preliminary,
Eq. (\ref{ratio}) demonstrates the requirement for a high mechanical quality factor, which is selected as $10^6$ in Fig. \ref{fig3}.
Otherwise, we cannot obtain a small value for $n_{0,opt}$, which is necessary for ground-state cooling.

When $\kappa/\omega_1$ and $\omega_1/\gamma_1$ are fixed, Eq. (\ref{ratio})
also provides indications of the requirements for pre-cooling (which determines $n_{1,th}$)
and the effective coupling strength $g_1 \alpha_s$.
Meanwhile, the
value of $\omega_0+2\Delta$ in the Eq. (\ref{ratio}) denominator cannot be too small.

\subsubsection{Selection of optimal detuning}\label{detuning}

The factor $1/N[\omega]$ is absent from Eq. (\ref{ratio}), but
must be taken into account when comparing the scales of 
the two terms in the first line of Eq. (\ref{2-sources}).
$N[\omega_0]$ of small magnitude could significantly enhance the value
of $S_{nn}[\omega_0]$, and also result in
a large $\gamma_{opt,0}$.
The remaining term $\chi_{c,\omega}$ is sufficiently small in magnitude to be neglected.
From Eq. (\ref{approximations}), it is apparent that 
\begin{equation}
|N[\omega]|=|\omega_0^2-\tilde{\omega}_1^2|,
\end{equation}
near $\omega=\pm\omega_0$. Thus, the condition $|\omega_0-\tilde{\omega}_1|\ll\omega_{0/1}$
proposed in the previous section is satisfied.
According to Eq. (\ref{tilde-w1}), $\omega_1-\tilde{\omega}_1\approx6 g_1^2\alpha_s^2\Delta$. Therefore, in the optimal
cooling region, $g_1\alpha_s$ and $\Delta$ exhibit an inverse relationship.
We will verify this point via numerical calculation in the next section.

The small magnitude of $N[\omega]$ also amplifies its potential
contribution to $S_{nn}[-\omega_0]$. Fortunately, this contribution vanishes if
$$\alpha(-\omega_0)=0\qquad\Rightarrow\qquad \Delta\approx\frac{\omega_0}{2}.$$
This kind of destructive interference is inherited from the
dissipative optomechanical system (see Sec. \ref{intro-diss}). 

The statements made in this subsection are verified in Fig. \ref{fig3}. This image
shows that, when $\Delta=\omega_0/2$, the primary contribution to the left peak is
from the thermal environment (blue dashed line), whereas the right
peak is primarily due to the photon number fluctuation.

\begin{figure*}[t]
\includegraphics[width=14cm]{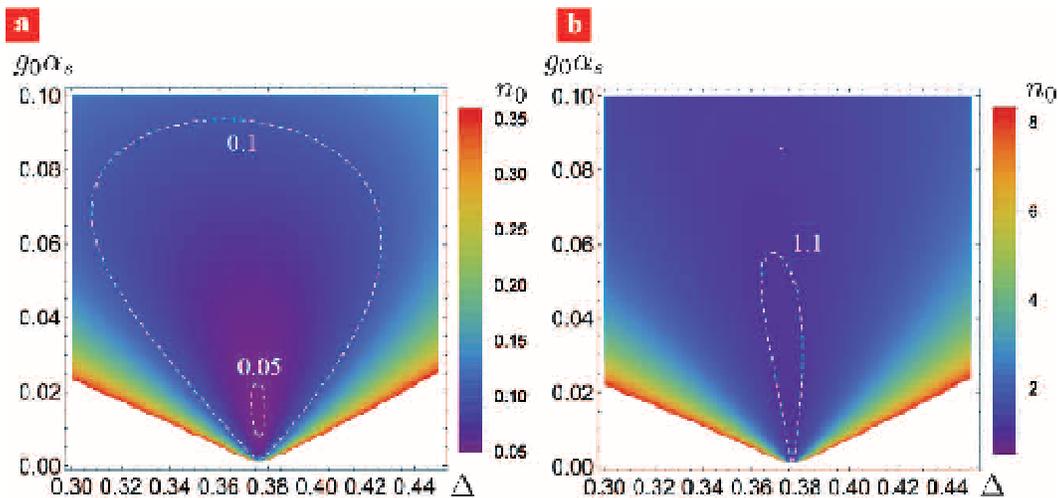}
\caption{Cooling limit $n_0$ for oscillator 0. $\omega_1$
is set as the unit, with $\gamma_0=\gamma_1=10^{-6}$,
$g_1\alpha_s=0.336$, $n_{0,th}=n_{1,th}=100$, and $\omega_0=0.7$. a) $\kappa=300$, such that $\kappa/\omega_0\approx 430$.
In the region enclosed by the two white dashed circles, $\tilde{n}_0 < 0.1$
and 0.05, as indicated in the figure. b) $\kappa=7000$,
such that $\kappa/\omega_0=10^{4}$, and $n_0 < 1.1$
in the region enclosed by the dashed line.
In the uncolored regions, $n_0$ is too high to show.}
\label{fig4}
\end{figure*}

\section{Ground-state cooling: exact phonon-number solutions}\label{exact-section}

In the previous section, we determined the possibility of
ground-state cooling using quantum noise analysis.
In this section, we shall examine this further by solving the
linearized quantum Langevin equations exactly, i.e,
Eqs. (\ref{equ-d})--(\ref{equ-b1}).
Note that the expression for $\hat{b}_{0,\omega}$ is presented in the Appendix, as it is very complex.
The exact result of the cooling limit $n_0$ is obtained from the integration
\begin{equation}
n_0=\int_{-\infty}^{\infty}\frac{d\omega}{2\pi}\frac{d\omega'}{2\pi}\langle\hat{b}^\dagger_{0,\omega}
\hat{b}_{0,\omega'}\rangle.
\label{n0}
\end{equation}

\subsection{Cooling in the unresolved sideband regime}
Firstly, we examine the prediction of ground-state cooling in the unresolved sideband regime.
The dependence of $n_0$ on $g_0\alpha_s$ and $\Delta$ is illustrated in Fig. (\ref{fig4}).
This figure shows that the phonon number of the target oscillator $\tilde{n}_0$ can be reduced to less than $0.05$ when
the sideband parameter $\kappa/\omega_0$ is approximately $430$.
Moreover, when $\kappa/\omega_0$ is increased to $10^4$,
$n_0\approx 1$ can still be decreased to approximately 1.
(In comparison, the standard sideband cooling at $\Delta=-\kappa/2$ yields
the optimal $n_{opt,0}=2500$.)

\subsection{Precision of quantum noise approach}

The quantum noise approach is based on the Fermi golden rule, which is valid only
for the weak limit.
Fig. \ref{fig4} also shows the region that is invisible to the quantum noise approach.
This figure indicates that, when $g_1$ is fixed, optimal cooling is realized with proper $g_0$.
However, the quantum noise approach naively suggests larger $g_0$.

\begin{figure}[b]
\includegraphics[width=0.47\textwidth]{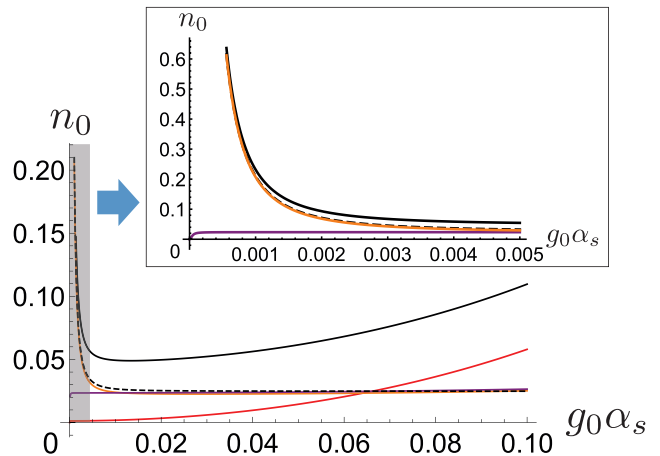}
\caption{$n_0$ as function of $g_0$ (and magnified for weak couplings).
$\omega_1$ is set as the unit,
$\Delta=0.377$, and the other parameters are identical to those in Fig. \ref{fig3}.
The black dashed line is the phonon occupation
given by the quantum noise approach, the black solid line is the
exact result for $n_0$, which is the sum of the three contributions
from the laser driving fluctuation $\hat{d}_{in}$ (red), the
local thermal environment $\hat{b}_{0,in}$ (orange), and
the thermal environment of the ancillary oscillator $\hat{b}_{1,in}$ (purple).}
\label{fig5}
\end{figure}

To visualize the breakdown of the quantum noise approach, we
note that $n_0$ is determined independently in terms of the laser-driving fluctuation noise,
the local thermal environment noise, and the thermal environment of the ancillary oscillator.
In Fig. \ref{fig5}, these contributions are illustrated separately.
We find that the laser-driving
fluctuation dominates the phonon source after $n_0$ has been obtained,
as shown by the red line in Fig. \ref{fig5}.
As the source of this contribution is independent of
the thermal environment of the oscillators, it provides an intrinsic limit for the optically induced cooling.

The red line in Fig. \ref{fig5}, or the term for the photon number fluctuation noise,
is ascribed to a term in the
exact solution of $\hat{b}_{0,\omega}$ [Eq. (\ref{phononb}) in the Appendix]:
\begin{equation}
\begin{aligned}
&\frac{-i g_0\alpha_s \sqrt{\kappa}\chi^{-1*}_{0,-\omega}}{\mathcal{N}[\omega]}
\{ \chi^{-1}_{c,\omega}\chi^{-1}_{1,\omega}\chi^{-1*}_{1,-\omega} \\
   & \qquad +i 2 \omega_1 g_1^2\alpha_s^2 (2\Delta^2-i\Delta \kappa-i\omega \kappa)\}\hat{d}^\dagger_{in,\omega}.
\end{aligned}
\end{equation}
Here, $\mathcal{N}[\omega]$ is given as
\begin{equation}
\begin{aligned}
\mathcal{N}[\omega]= & \chi^{-1}_{c,\omega}\chi^{-1*}_{c,-\omega}\chi^{-1}_{0,\omega}\chi^{-1*}_{0,-\omega}N[\omega]\\
& -4\omega_0\omega_1\kappa^2g_0^2g_1^2\alpha_s^4 + 4\Delta\omega_0
 g_0^2\alpha_s^2\chi^{-1}_{1,\omega}\chi^{-1*}_{1,-\omega}.
\end{aligned}
\end{equation}
Further, $N[\omega]$ has already been given in Eq. (\ref{near}).
$1/\mathcal{N}[\omega]$ is actually the common factor
of all the other terms involving $\hat{d}_{in}$, $\hat{b}_{0,in}$, and so on, and
its poles determine the eigen-frequencies and damping rates
of all the eigenmodes of the hybrid system.

The quantum noise approach
functions well in the weak regime. In our derivation of $S_{nn}[\omega]$, we treat the optical mode
and oscillator 1 as the steady background; this approach is valid if their dynamics
are significantly faster than that of oscillator 0.
Because we have $\kappa\gg\tilde{\gamma}_1\gg\gamma_0$, the quantum noise approach
is effective, as verified by the magnified chart shown in Fig. \ref{fig5}.

It is emphasized that, in Fig. \ref{fig5}, there is a coincidence between the quantum noise predictions and
the $n_0$ component contributed by the local environment, i.e., the orange lines
obtained from the $\langle\hat{b}_{in,0,\omega}\hat{b}_{in,0,\omega'}^\dagger\rangle$
and $\langle\hat{b}_{0,in,\omega}^\dagger\hat{b}_{0,in,\omega'}\rangle$ terms in Eq. (\ref{n0}).
This is reasonable, because only
the $n_0$ contributions from the local noise decrease with increased $g_0$ (at least in the plotted region).

\subsection{Inverse relationship between $g_1\alpha_s$ and optimal detuning}

\begin{figure}[t]
\includegraphics[width=0.4\textwidth]{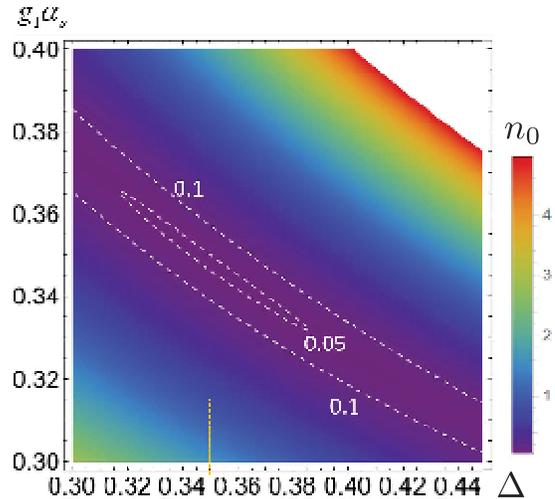}
    \caption{Mean phonon number $n_0$ as a function of $g_1\alpha_s$ and the detuning $\Delta$. $\omega_1$
    is set as the unit, with $\gamma_0=\gamma_1=10^{-6}$,
    $g_0\alpha_s=0.1$, $n_{0,th}=n_{1,th}=100$, and $\omega_0=0.7$. $\kappa=300$, such that $\kappa/\omega_0\approx 430$.
    In the region enclosed by the white dashed lines, $\tilde{n}_0 < 0.1$ and $0.05$, as indicated in the figure.
    The position of $\Delta=\omega_0/2$ is indicated by a short dashed line.}
  \label{fig6}
\end{figure}

In Sec. \ref{detuning}, we have shown that the quantum noise approach suggests $\Delta\approx\omega_0/2$,
along with an inverse relationship
between $g_1\alpha_s$ and $\Delta$.
Here, we examine the quantum noise approach using the exact solution of $n_0$ and illustrate the result in Fig. (\ref{fig6}).
The numerical result supports this statement very well.
Additionally, it can be seen from Fig. (\ref{fig4}) that
the optimal detuning exhibits independence of the sideband parameter when $g_0\alpha_s$ is very close to zero.
This feature also agrees with the theory presented in Sec. \ref{detuning}.

\section{Discussion}\label{discussion}
A remarkable proposal has demonstrated that strong and tunable dissipative
coupling can be established by inserting a movable membrane in a Michelson-Sagnac
interferometer~\cite{diss-realize}. Hence, a hybrid system of dispersively and dissipatively
coupled oscillators can be constructed by replacing one fixed mirror within the interferometer with a movable one.
Our scheme requires only a single optical mode with one driving mode.
Therefore, in cases where more dissipatively coupled oscillators are accessible,
our approach is significantly simpler than schemes involving ground-state cold atom ensembles or
correlated multi-cavities. It is also simpler than those based on OMIT
effects~\cite{eit-2,eit-cooling}, where two driving modes and an additional
oscillator in the resolved sideband regime are required.
From a practical perspective, the most important concern is successfully embedding a dissipative oscillator
into other dispersive optomechanical systems to obtain ground-state cooling.
We do not elaborate on this topic here, because only a small number of experimental demonstrations
of dissipative optomechanical systems have been reported.

In the following, we discuss commonalities and conduct a comparison between the approach presented in this paper and the existing unresolved sideband cooling schemes.
First, there are similarities between the Hamiltonians used in the various cooling schemes and that proposed here,
characterized by $H_{int,1}$ [Eq. (\ref{dissi-couple})].
In the OMIT scheme, the Hamiltonian \cite[Eq. (49)]{eit-cooling} is
\begin{equation}
H_{int}=\hbar g_2(\hat{b}_2^\dagger+\hat{b}_2)(\hat{a}^\dagger_0\hat{a}_1+\hat{a}^\dagger_1\hat{a}_0),
\end{equation}
where $\hat{a}_{0/1}$ are the two required cavity modes.
In the atom-optomechanical hybrid system, the Hamiltonian \cite[Eq. (3)]{atom-2015} is
\begin{equation}
H_{int}=G_0(\hat{a}^\dagger\hat{c}+\hat{c}^\dagger\hat{a}),
\end{equation}
where $\hat{a}$ and $\hat{c}$ correspond to the optical mode and collective operators of the atom ensemble, respectively.
Finally, in the coupled-cavity scheme, the Hamiltonian~\cite[Eq. (A2)]{mt-cav-2015} is
\begin{equation}
H_{int}=J\hat{a}_1^\dagger\hat{a}_2+J^*\hat{a}_2^\dagger\hat{a}_1,
\end{equation}
where $\hat{a}_1$ is the optical mode coupled with the mechanical oscillator
and $\hat{a}_2$ is the mode of the ancillary cavity, the line width of which must be significantly smaller than
the resonant frequency of the target oscillator.
It is quite interesting to note the similarities between these damping-like interactions.
From this perspective, the ancillary oscillator
serves as a medium in our scheme, and it is the input/environment mode that is essential for
ground-state cooling. The ancillary oscillator plays a role similar to those of the
ancillary cavities or ground-state atoms in the alternative schemes.

Next, we compare our scheme more closely with the OMIT
scheme \cite{eit-cooling,eit-2}, which is analogous to electromagnetically induced
transparency. The OMIT scheme uses an ancillary oscillator that is
dispersively coupled to the cavity mode. Its position spectrum is also embedded in the
photon-number fluctuation spectrum.
In addition, a significant imbalance between the two
peaks of the position spectrum is necessary.
Therefore, the ancillary dispersive oscillator must also be cooled to the ground state.
However, as this oscillator is coupled dispersively, cooling is only possible if the resonant frequency
is larger than the cavity line width.
This leads to an extremely large gap between the frequencies of the two oscillators.
In our notation, this large frequency gap means that the requirement $\omega_0\approx \tilde{\omega}_1$ cannot be satisfied.
Thus, another largely detuned driving laser would be required in order to introduce the driving laser beat note, and
unexpected and harmful excitation
of some modes of the real mechanical system could be induced.
(see \cite[Sec. \uppercase\expandafter{\romannumeral3}.G.]{eit-cooling}).
The method using the dissipative oscillator presented in this study avoids this problem.

\section{Conclusion and Outlook}\label{conclusion}
In this paper, we have proposed a new scheme for optomechanical
cooling in the unresolved sideband regime,
and we have verified this technique using both the quantum noise approach and exact solutions to
the quantum Langevin equations.
The proposed scheme uses a dispersively coupled oscillator that
is also in the unresolved sideband regime. This ancillary oscillator significantly
modifies the photon-number fluctuation spectrum and, thus, realizes
ground-state cooling in the unresolved sideband regime.
Therefore, the dissipatively coupled oscillator can cool not only itself, but also other
mechanical oscillators coupled with the same optical mode.
This scheme will enrich the optomechanical toolbox. Further, as dissipatively
coupled systems have not been investigated widely, this result will stimulate further
interest and related research questions.

As optomechanics facilitates the realization of controllable macroscopic quantum systems, 
it is feasible that this subject will play
an increasingly important role in quantum metrology, quantum information processing,
and, possibly, in future quantum computers.
The very recent confirmation of gravitational waves using optomechanical technology \cite{gwave} has strongly
endorsed this confidence.
The scheme developed in this article will enrich the optomechanical cooling toolbox. 
In addition, the use of dissipative coupling in cooling constitutes an example of beneficial exploitation of the noisy environment modes.
Finally, as dissipatively coupled systems have not been investigated widely, 
we hope that this result will stimulate further interest in this topic, along with related research questions.

\acknowledgments
The authors thank Haixing Miao for useful discussion.
This work was supported by the Joint Studies Program of the Institute for Molecular Science,
the NINS Youth Collaborative Project, JSPS KAKENHI (Grant No. 25800181), the DAIKO Foundation,
the National Natural Science Foundation of China (Grants Nos. 11275181 and 61125502),
the National Fundamental Research Program of China (Grant No. 2011CB921300), and the Chinese Academy of Science (CAS).
Y.-X. Z. is grateful for the hospitality of the Institute for Molecular Science, National Institutes of National Sciences,
under the IMS International Internship program.

\appendix
\section{Exact solution of phonon annihilation operator}
The phonon annihilation operator $\hat{b}_{0,\omega}$ is expressed exactly as
\begin{equation}
\begin{aligned}
\hat{b}_{0,\omega}= & \frac{-ig_0\alpha_s \sqrt{\kappa}\chi^{-1*}_{0,-\omega}}{\mathcal{N}[\omega]}
\{\chi^{-1*}_{c,-\omega}\chi^{-1}_{1,\omega}\chi^{-1*}_{1,-\omega}\\
& \qquad +4i\omega_1 g_1^2\alpha_s^2\Delta(i\kappa-\Delta)\}\hat{d}_{in,\omega}\\
&+\frac{-i g_0\alpha_s \sqrt{\kappa}\chi^{-1*}_{0,-\omega}}{\mathcal{N}[\omega]}
\{ \chi^{-1}_{c,\omega}\chi^{-1}_{1,\omega}\chi^{-1*}_{1,-\omega} \\
   & \qquad +i 2 \omega_1 g_1^2\alpha_s^2 (2\Delta^2-i\Delta \kappa-i\omega \kappa)\}\hat{d}^\dagger_{in,\omega}\\
& -\frac{\sqrt{\gamma_0}}{\mathcal{N}[\omega]}\{2ig_0^2\alpha_s^2(\kappa^2\omega_1g_1^2\alpha_s^2-
\Delta\chi^{-1}_{1,\omega}\chi^{-1*}_{1,-\omega})\\
& \qquad +\chi^{-1}_{c,\omega}\chi^{-1*}_{c,-\omega}\chi^{-1*}_{0,-\omega}N[\omega]\}\hat{b}_{in,0,\omega}\\
& +\frac{2i\sqrt{\gamma_0}g_0^2\alpha_s^2}{\mathcal{N}[\omega]}(\Delta \chi^{-1}_{1,\omega}
\chi^{-1*}_{1,-\omega}-\omega_1\kappa^2g_1^2\alpha_s^2) \hat{b}_{in,0,\omega}^\dagger\\
& +\frac{g_0^2 g_1^2\alpha_s^4\chi^{-1*}_{0,-\omega}\sqrt{\gamma_1}}{2\mathcal{N}[\omega]}
(i\kappa^2+2\kappa\omega-4i\Delta^2)\\
&\qquad \times (\chi^{-1*}_{1,-\omega} \hat{b}_{in,1,\omega} + \chi_{1,\omega}^{-1}\hat{b}^\dagger_{in,1,\omega}).
\end{aligned}
\label{phononb}
\end{equation}

\end{document}